# Three-level Λ-type atomic systems with a Pseudo-Hermitian PT-symmetric Hamiltonian

Amarendra K. Sarma* and Balla Prannay
Department of Physics, Indian Institute of Technology Guwahati, Guwahati-781039, Assam, India
*Electronic address: aksarma@iitg.ernet.in

We have studied a three-level Λ-type atomic system with all the energy levels exhibiting decay. The system is described by a pseudo-Hermitian Hamiltonian and subject to certain conditions, the Hamiltonian shows parity-time (PT) symmetry. The probability amplitudes of various atomic levels both below and above the PT-theshold is worked out.

**PACS** number(s)**:** 11.30.Er, 03.65.Ge, 42.82.Et

Since the pioneering work of Bender and Boettcher [1], the fundamental importance of non-Hermitian operators obeying pure real spectra is widely recognized and studied [2-6]. Among this wide class of non-Hermitian Hamiltonians, there exists some Hamiltonians that respects the parity-time (PT) symmetry. In general, for the eigenvalues of the non-Hermitian Hamiltonian $\widehat{H} = \frac{\hat{p}^2}{2m} + V(\hat{x})$, where $m$ is the mass, $\hat{x}$ is the position operator, $\hat{p}$ is the momentum operator, $V$ is the complex potential, to be entirely real, the Hamiltonian should share common set of eigen-vectors with the PT-operator such that $[H, PT] = 0$. $P$ is the parity inversion operator defined as: $\hat{p} \to -\hat{p}, \hat{x} \to -\hat{x}$ and $T$ is the time-reversal operator which is defined as follows: $\hat{p} \to -\hat{p}, \hat{x} \to -\hat{x}, i \to -i$ [1]. It turns out that the Hamiltonian will be PT-symmetric only when $V^*(-x) = V(x)$. It is remarkable to note that the idea of parity-time symmetry is explored in as diverse areas as quantum field theory, open quantum systems, metamaterials, optical physics, electronics and plasmonics [2-13]. In fact, after the experimental demonstration of parity-time symmetry in optics and other areas [7-11], studies related to PT-symmetric non-Hermitian systems are getting a tremendous boost [14-20]. Recently, PT-symmetry is explored in coherently prepared three and four level atomic systems [21,22]. In all such studies, generally damping of the ground state is ignored. In this short communication, we report a study on a three-level atomic system with damping effects. The system is described by a pseudo-Hermitian Hamiltonian. A Hamiltonian, $H$, is termed as pseudo-Hermitian if [6]: $\eta H \eta^{-1} = H^\dagger$, where $\eta$ is called the metric and is developed through the definition of distorted inner product $< \psi | \eta \psi >$ [23-25].

The schematic of our model is shown in Fig. 1. We consider a Λ–type atomic system. Each of the two laser pulses interacts only with a pair of energy levels. The pump ($E_p$) and coupling ($E_c$) fields drive the atomic transitions $|1> - |2>$ and $|2> - |3>$, respectively. The transition $|1> - |3>$ is dipole forbidden. The atomic system is assumed to be in a superposition of eigenstates given by:

$$|\psi> = C_1|\varphi_1> + C_2|\varphi_2> + C_3|\varphi_3> \qquad (1)$$

Here $C_1$, $C_2$ and $C_2$ are the time dependent probability amplitudes corresponding to the levels |1>, |2> and |3> respectively.

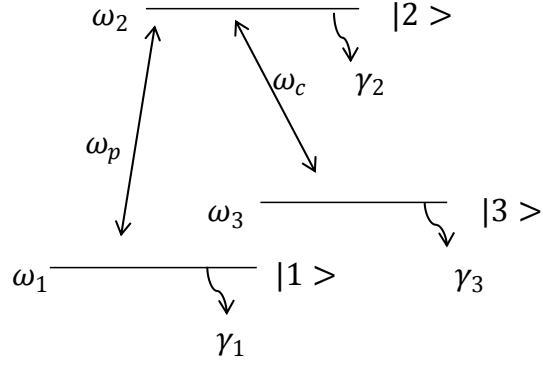

FIG. 1 Schematic of the 3 level atomic systems under consideration

Damping of the atoms are described by the phenomenological decay constants $\gamma_1$, $\gamma_2$ and $\gamma_3$ for the levels |1>, |2> and |3> respectively. The evolution of the probability amplitudes, following the so called Schrodinger equation, is given by:

$$i\hbar \frac{\partial}{\partial t}\begin{bmatrix}C_1\\C_2\\C_3\end{bmatrix} = \begin{bmatrix}-i\gamma_1 + \hbar\omega_1 & V_{12} & 0\\ V_{21} & -i\gamma_2 + \hbar\omega_2 & V_{23}\\ 0 & V_{32} & -i\gamma_3 + \hbar\omega_3\end{bmatrix}\begin{bmatrix}C_1\\C_2\\C_3\end{bmatrix} \qquad (2)$$

Here $V_{ij} = -e\vec{x_{ij}}\cdot\vec{E}$ are the matrix elements of the atom-radiation interaction potential with the total electric field, $\vec{E}$, given by: $\vec{E} = \frac{1}{2}\hat{\mathcal{E}}_c E_c e^{-i\omega_c t} + \frac{1}{2}\hat{\mathcal{E}}_p E_p e^{-i\omega_p t} + c.c.$ and $\omega_k$ is the eigen frequency of the state $|k>$. Now, we use the following transformation in Eq. (2):

$$\begin{bmatrix}C_1\\C_2\\C_3\end{bmatrix} = \begin{bmatrix}a_1\\a_2\\a_3\end{bmatrix}\exp[\gamma_2 t - i\omega_1 t] \qquad (3),$$

where we assume $\gamma_2 = (\gamma_1 + \gamma_3)/2$ and then taking the ansatz:

$$\begin{bmatrix}b_1(t)\\b_2(t)\\b_3(t)\end{bmatrix} = \begin{bmatrix}a_1(t)\\a_2(t)\exp(i\omega_p t)\\a_3(t)\exp\{i(\omega_p - \omega_c)t\}\end{bmatrix} \qquad (4),$$

we obtain, after a bit of simple algebra, the following evolution equations:

$$i\hbar\frac{\partial}{\partial t}\begin{bmatrix}b_1(t)\\b_2(t)\\b_3(t)\end{bmatrix} = \begin{bmatrix}-i\gamma_{pt} & V_{21}^{*p} & 0\\ V_{21}^p & \hbar\Delta & V_{23}^c\\ 0 & V_{23}^{*c} & +i\gamma_{pt} + \hbar\Delta\end{bmatrix}\begin{bmatrix}b_1(t)\\b_2(t)\\b_3(t)\end{bmatrix} = H\begin{bmatrix}b_1(t)\\b_2(t)\\b_3(t)\end{bmatrix} \qquad (5)$$

Here, $\gamma_{pt} = (\gamma_1 - \gamma_3)/2$; $V_{ij}^k = -e\vec{x_{ij}}\cdot\hat{\mathcal{E}}_k E_k/2$ (with i, j=1,2,3 and $k = c,p$) and $\Delta = \omega_{21} - \omega_p$. In obtaining Eq. (5), we assume, $\omega_{23} = \omega_c$. In these equations, $\omega_{ij} = \omega_i - \omega_j$. The above Hamiltonian is PT-symmetric, i.e $[H, PT] = 0$, subject to the following conditions: $V_{21}^p = V_{23}^c$ and $\Delta = 0$. It is assumed that $V_{21}^p$ and $V_{23}^c$ are both real. The resulting PT symmetric-Hamiltonian can now be written as:

$$H = \begin{bmatrix}-i\gamma_{pt} & V_{21}^p & 0\\ V_{21}^p & 0 & V_{21}^p\\ 0 & V_{21}^p & +i\gamma_{pt}\end{bmatrix} \qquad (6)$$

It should be noted that the above Hamiltonian is not Hermitian because of the phenomenological decay terms. The Hamiltonian in Eq. (6) is diagonalizable with the following eigenvalues:

$$E_0 = 0, E_\pm = \pm \sqrt{2V_{21}{}^{p^2} - \gamma_{pt}{}^2} \qquad (7)$$

The eigenvalues are real if, $2V_{21}{}^{p^2} > \gamma_{pt}{}^2$. The similarity matrix associated with diagonalization is found to be:

$$D = \frac{1}{E}\begin{bmatrix} \frac{(E-i\gamma_{pt})^2}{2V_{21}{}^p} & \frac{(E+i\gamma_{pt})^2}{2V_{21}{}^p} & -V_{21}{}^p \\ E - i\gamma_{pt} & -(E + i\gamma_{pt}) & -i\gamma_{pt} \\ V_{21}{}^p & V_{21}{}^p & V_{21}{}^p \end{bmatrix} \qquad (8)$$

The three eigenvectors are given by:

$$|\phi_1> = \frac{1}{E}\begin{bmatrix} \frac{(E-i\gamma_{pt})^2}{2V_{21}{}^p} \\ E - i\gamma_{pt} \\ V_{21}{}^p \end{bmatrix}, |\phi_2> = \frac{1}{E}\begin{bmatrix} \frac{(E+i\gamma_{pt})^2}{2V_{21}{}^p} \\ -(E + i\gamma_{pt}) \\ V_{21}{}^p \end{bmatrix}, |\phi_3> = \frac{1}{E}\begin{bmatrix} -V_{21}{}^p \\ -i\gamma_{pt} \\ V_{21}{}^p \end{bmatrix} \qquad (9)$$

We now construct the positive definite metric, $\eta$, by using $\eta = (DD^\dagger)^{-1}$ [25]:

$$\eta = \frac{1}{2E^2}\begin{bmatrix} 3V_{21}{}^{p^2} & 3i\gamma_{pt}V_{21}{}^p & -(V_{21}{}^{p^2} + \gamma_{pt}{}^2) \\ -3i\gamma_{pt}V_{21}{}^p & 2(V_{21}{}^{p^2} + \gamma_{pt}{}^2) & 3i\gamma_{pt}V_{21}{}^p \\ -(V_{21}{}^{p^2} + \gamma_{pt}{}^2) & -3i\gamma_{pt}V_{21}{}^p & 3V_{21}{}^{p^2} \end{bmatrix} \qquad (10)$$

It should be noted that it is possible to choose a different diagonalization matrix and consequently a different metric $\eta$. As a consequence of the metric $\eta$, we get $<\phi_m|\eta\phi_n> = \delta_{mn}$. The time evolution of the state describing the three-level atomic system can be expressed as:

$$|\psi> = C_1''(t)|\phi_1> + C_2''(t)|\phi_2> + C_3''(t)|\phi_3> \qquad (11)$$

Here $C_1''(t) = C_1''(0)e^{\frac{-iEt}{\hbar}}$, $C_2''(t) = C_2''(0)e^{\frac{+iEt}{\hbar}}$ and $C_3''(t) = C_3''(0)$.

The coefficients $C_i''(t)(i = 1,2,3)$ can be written in terms of $b_1(t)$, $b_2(t)$ and $b_3(t)$ using the following equation:

$$\begin{bmatrix} b_1(t) \\ b_2(t) \\ b_3(t) \end{bmatrix} = D \begin{bmatrix} C_1''(t) \\ C_2''(t) \\ C_3''(t) \end{bmatrix} \qquad (12)$$

One can find the coefficients $C_i''(t)$ by simply taking the inverse transform of Eq. (12). We can easily calculate the coefficients as follows:

$$b_1(t) = \frac{V_{21}{}^p}{E}(C_1''(t) + C_2''(t) - C_3''(t)) - \frac{i\gamma_{pt}}{V_{21}{}^p}(C_1''(t) - C_2''(t)) - \frac{\gamma_{pt}{}^2}{V_{21}{}^p E}(C_1''(t) + C_2''(t)) \qquad (13a)$$

$$b_2(t) = C_1''(t) - C_2''(t) - \frac{i\gamma_{pt}}{E}(C_1''(t) + C_2''(t) + C_3''(t)) \qquad (13b)$$

$$b_3(t) = \frac{V_{21}{}^p}{E}(C_1''(t) + C_2''(t) + C_3''(t)) \qquad (13c)$$

It may be useful and instructive to look at a specific case of the three-level atomic system with real eigenvalues, say the atom is initially at its lowest energy level, i.e. $b_1(0) = 1, b_2(0) = 0$ and $b_3(0) = 0$. We obtain, after some algebra:

$$b_1(t) = \frac{V_{21}{}^{p^2}}{E^2}\left(\cos\frac{Et}{\hbar} + 1\right) + \frac{i\gamma_{pt}}{E}\left(\sin\frac{Et}{\hbar}\right) - \frac{\gamma_{pt}{}^2}{E^2}\left(\sin\frac{Et}{\hbar}\right) \qquad (14a)$$

$$b_2(t) = -\frac{iV_{21}{}^p}{E}\left(\sin\frac{Et}{\hbar}\right) - \frac{i\gamma_{pt}V_{21}{}^p}{E^2}\left(\cos\frac{Et}{\hbar} - 1\right) \tag{14b}$$

$$b_3(t) = \frac{V_{21}{}^{p2}}{E^2}\left(\cos\frac{Et}{\hbar} - 1\right) \tag{14c}$$

The original coefficients could be easily worked out from Eq. (3) and (4). They are given below:

$$C_1 = \left(\frac{V_{21}{}^{p2}}{E^2}\left(\cos\frac{Et}{\hbar} + 1\right) + \frac{i\gamma_{pt}}{E}\left(\sin\frac{Et}{\hbar}\right) - \frac{\gamma_{pt}{}^2}{E^2}\left(\sin\frac{Et}{\hbar}\right)\right) e^{-\frac{\gamma_2 t}{\hbar} + i\omega_1 t} \tag{15a}$$

$$C_2 = \left(-\frac{iV_{21}{}^p}{E}\left(\sin\frac{Et}{\hbar}\right) - \frac{i\gamma_{pt}V_{21}{}^p}{E^2}\left(\cos\frac{Et}{\hbar} - 1\right)\right) e^{-\frac{\gamma_2 t}{\hbar} + i(\omega_1 - \omega_p)t} \tag{15b}$$

$$C_3 = \frac{V_{21}{}^{p2}}{E^2}\left(\cos\frac{Et}{\hbar} - 1\right) e^{-\frac{\gamma_2 t}{\hbar} + i(\omega_1 + \omega_c - \omega_p)t} \tag{15c}$$

It may be noted that the normalization condition, $|C_1(t)|^2 + |C_2(t)|^2 + |C_3(t)|^2 = 1$, is not met here owing to the presence of the damping terms. It is easy to see from Eq. (15) that subject to the condition, $2V_{21}{}^{p2} > \gamma_{pt}{}^2$, the atomic system decays by the average decay constant $\gamma_2 = (\gamma_1 + \gamma_3)/2$, provided the system obeys $2V_{21}{}^{p2} > \gamma_{pt}{}^2$. The coefficients oscillate with a frequency $\sqrt{2V_{21}{}^{p2} - \gamma_{pt}{}^2}$ and it depends on the strength of the pulses and the coherence between the states. In order to have a qualitative understanding of the populations of various levels, in Fig. 2 we depict the time evolution of the populations in the three energy levels of the atomic system for two sets of parameters, explained in the figure captions.

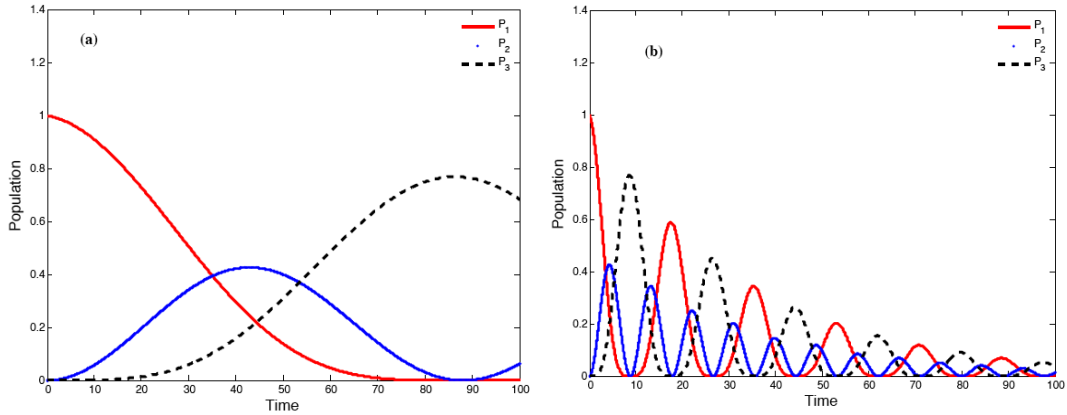

Fig. 2 (Color online) Time evolution of populations with (a) $\gamma_1 = .002, \gamma_2 = .0015, \gamma_3 = .001$ and $V = 0.025$ (b) $\gamma_1 = .02, \gamma_2 = .015, \gamma_3 = .01$ and $V = 0.25$. We put $\hbar = 1$.

We observe that populations oscillate between the atomic states. However, if the dampings of the levels are increased populations in various levels decay, and this decay rate increases with increase in the damping coefficients. Also we find that the middle energy level should have comparatively higher life-time than that of the other two levels.

It may be straightforward to check the case of broken PT-symmetry, i.e. when, $2V_{21}{}^{p2} < \gamma_{pt}{}^2$. In this case two of the eigenvalues are purely imaginary. The Hamiltonian still remains to be pseudo-Hermitian. Thus, we can construct a metric and carry out similar analysis. Our calculations lead to the following expressions for the coefficients:

$$C_1(t) = \left[\frac{-V_{21}p^2}{\gamma_{pt}^2 - 2V_{21}p^2}\left(\cosh\left(\frac{\sqrt{\gamma_{pt}^2 - 2V_{21}p^2}\,t}{\hbar}\right) + 1\right) - \frac{\gamma_{pt}}{\sqrt{\gamma_{pt}^2 - 2V_{21}p^2}}\sinh\left(\frac{\sqrt{\gamma_{pt}^2 - 2V_{21}p^2}\,t}{\hbar}\right) + \frac{\gamma_{pt}^2}{\gamma_{pt}^2 - 2V_{21}p^2}\sinh\left(\frac{\sqrt{\gamma_{pt}^2 - 2V_{21}p^2}\,t}{\hbar}\right)\right] e^{-\frac{\gamma_2 t}{\hbar} + i\omega_1 t} \quad (16a)$$

$$C_2(t) = \left[\frac{V_{21}p}{\sqrt{\gamma_{pt}^2 - 2V_{21}p^2}}\sinh\left(\frac{\sqrt{\gamma_{pt}^2 - 2V_{21}p^2}\,t}{\hbar}\right) + \frac{\gamma_{pt}V_{21}p}{\sqrt{\gamma_{pt}^2 - 2V_{21}p^2}}\left(\cosh\left(\frac{\sqrt{\gamma_{pt}^2 - 2V_{21}p^2}\,t}{\hbar}\right) - 1\right)\right] e^{-\frac{\gamma_2 t}{\hbar} + i(\omega_1 - \omega_p)t} \quad (16b)$$

$$C_3(t) = \frac{V_{21}p^2}{\sqrt{\gamma_{pt}^2 - 2V_{21}p^2}}\left(1 - \cosh\left(\frac{\sqrt{\gamma_{pt}^2 - 2V_{21}p^2}\,t}{\hbar}\right)\right) e^{-\frac{\gamma_2 t}{\hbar} + i(\omega_1 + \omega_c - \omega_p)t} \quad (16c)$$

In conclusion, we have studied a three-level Λ-type atomic system with damping terms. The system is described by a pseudo-Hermitian Hamiltonian. It is found that subject to certain conditions, the Hamiltonian is PT-symmetric. We have worked out the probability amplitudes of various levels both below and above the PT-theshold. The analysis may be applied to any real atom and the predictions of this work could be tested experimentally.

**Acknowledgements**
A.K.S. would like to acknowledge the financial support from DST, Government of India [Grant No. SB/FTP/PS-047/2013].